\documentclass[aps,twocolumn,prl,preprintnumbers,amsmath,amssymb,superscriptaddress,floatfix]{revtex4}

\usepackage{graphicx}% Include figure files
\usepackage{bm}% bold math
\usepackage{hyperref}
\usepackage{natbib}
\usepackage{braket}
\usepackage{float}

\begin{document}

\title{Singlet-triplet excitations and long range entanglement in the spin-orbital liquid candidate FeSc$_2$S$_4$}

\author{N. J. Laurita}
\affiliation{The Institute for Quantum Matter, Department of Physics and Astronomy, The Johns Hopkins University, Baltimore, MD 21218, USA}

\author{J. Deisenhofer}
\affiliation{Institute of Physics, University of Augsburg, 86135 Augsburg, Germany}

\author{LiDong Pan}
\affiliation{The Institute for Quantum Matter, Department of Physics and Astronomy, The Johns Hopkins University, Baltimore, MD 21218, USA}

\author{C. M. Morris}
\affiliation{The Institute for Quantum Matter, Department of Physics and Astronomy, The Johns Hopkins University, Baltimore, MD 21218, USA}

\author{M. Schmidt}
\affiliation{Experimental Physics~V, Center for Electronic
Correlations and Magnetism, University of Augsburg,
D-86135~Augsburg, Germany}

\author{M.~Johnsson}
\affiliation{Department of Materials and Environmental Chemistry, Stockholm University, 10691 Stockholm, Sweden}

\author{V.~Tsurkan}
\affiliation{Experimental Physics~V, Center for Electronic
Correlations and Magnetism, University of Augsburg,
D-86135~Augsburg, Germany} \affiliation{Institute of Applied
Physics, Academy of Sciences of Moldova, MD-2028~Chisinau, Republic
of Moldova}

\author{A. Loidl}
\affiliation{Experimental Physics~V, Center for Electronic
Correlations and Magnetism, University of Augsburg,
D-86135~Augsburg, Germany}

\author{N. P. Armitage}
\affiliation{The Institute for Quantum Matter, Department of Physics and Astronomy, The Johns Hopkins University, Baltimore, MD 21218, USA}

\date{\today}

\begin{abstract}
Theoretical models of the spin-orbital liquid (SOL) FeSc$_2$S$_4$ have predicted it to be in close proximity to a quantum critical point separating a spin-orbital liquid phase from a long-range ordered magnetic phase. Here, we examine the magnetic excitations of FeSc$_2$S$_4$ through time-domain terahertz spectroscopy under an applied magnetic field.  At low temperatures an excitation emerges that we attribute to a singlet-triplet excitation from the SOL ground state.  A three-fold splitting of this excitation is observed as a function of applied  magnetic field.  As singlet-triplet excitations are typically not allowed in pure spin systems, our results demonstrate the entangled spin and orbital character of singlet ground and triplet excited states.   Using experimentally obtained parameters we compare to existing theoretical models to determine FeSc$_2$S$_4$'s proximity to the quantum critical point.   In the context of these models, we estimate that the characteristic length of the singlet correlations to be  $\xi/ ( \textbf{a}/2) \approx 8.2$ (where  \textbf{a}/2 is the nearest neighbor lattice constant) which establishes FeSc$_2$S$_4$ as a SOL with long-range entanglement.

\end{abstract}

\maketitle

The search for ground states without classical analogs, e.g. quantum ground states, is a central focus of modern condensed matter physics.  A zero temperature spin liquid would be a prime realization of such a state \cite{Balents2010}.  Such systems possess local moments, but, due to quantum fluctuations, typically enhanced due to geometric frustration, do not order even at zero temperature.  They are proposed to have quantum mechanically entangled wavefunctions, with exotic fractionalized excitations.  Orbital degrees of freedom can also be disordered by quantum fluctuations \cite{Khaliullin2000}. Systems with both spin and orbital fluctuations as well as spin-orbit coupling (SOC) have been proposed to form a ``spin-orbital liquid" (SOL) ground state, characterized by entangled spin and orbital degrees of freedom but no long range order \cite{Fritsch2004,Chen2009a,Chen2009, Ole2012, Brzezicki2015}.

\begin{figure}[tb]
\includegraphics[width=1.0\columnwidth]{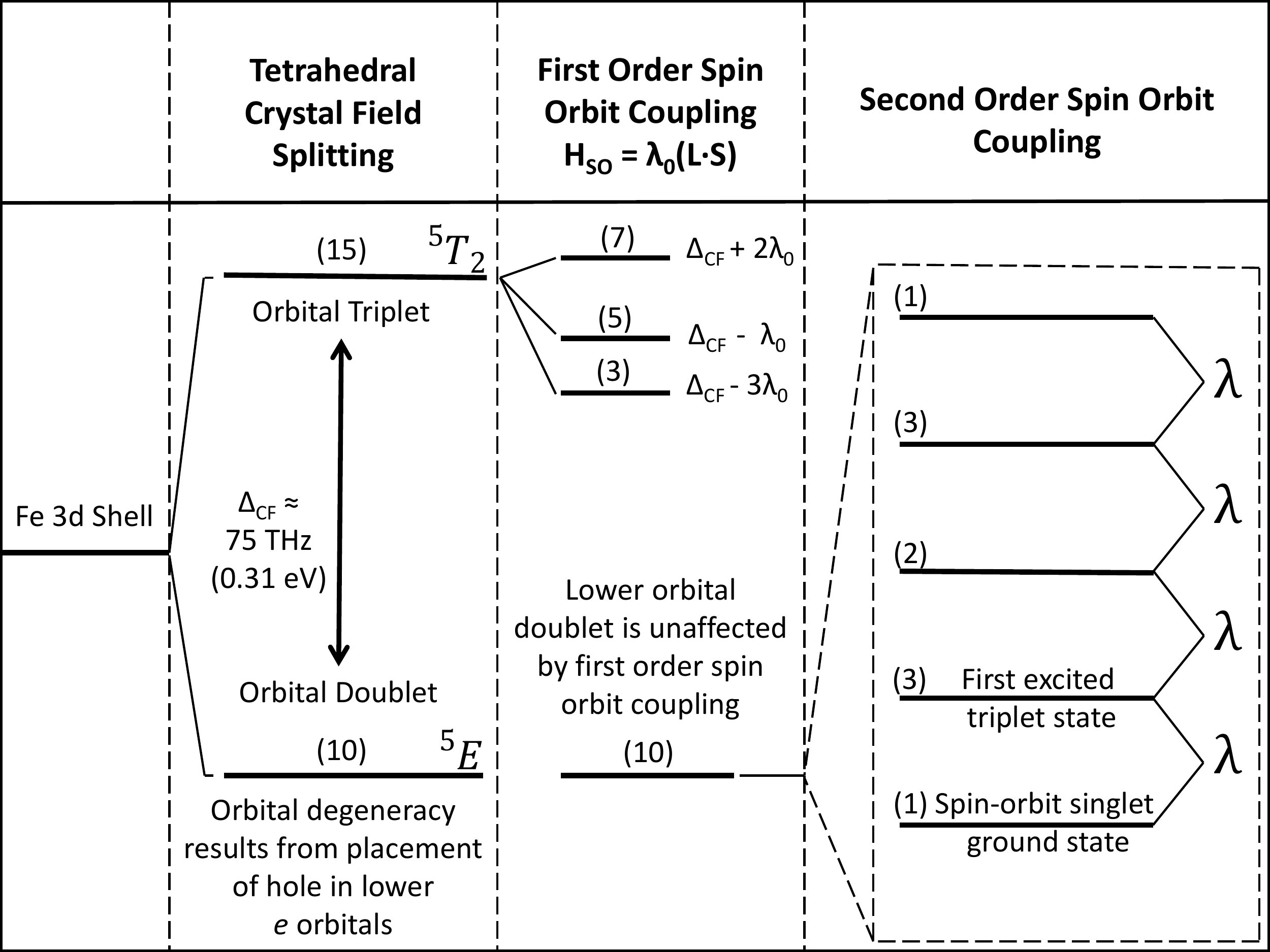}
\caption{(a) Energy levels of Fe$^{2+}$ ion in an S=2 configuration after tetrahedral crystal field splitting and first and second order SOC, without including lattice effects.  Numbers in parenthesis represent the degeneracy of each level.  Second order SOC splits the lower  $^5$E  doublet into 5 equally spaced levels separated by $\lambda = 6 ({\lambda _0}^2/\Delta _{CF})$.}
\label{Fig1}
\end{figure}

Recent experiments have shown a SOL phase may exist in the geometrically frustrated A site cubic spinel compound FeSc$_2$S$_4$.  A tetrahedral S$_4$ crystal field splits a $3d$ shell into an upper $t_{2}$ orbital triplet and a lower $e$ orbital doublet.  With Hund's coupling, an Fe$^{2+}$ ion in a tetrahedral environment assumes a high spin $S=2$ configuration with a lower $^5$E orbital doublet ground state and an upper $^5$T$_{2}$ orbital triplet excited state (Fig. \ref{Fig1}).  The ground state's two-fold orbital degeneracy is associated with the freedom to place a hole in either $e$ orbital.  Although orbital degeneracy is often relieved by Jahn-Teller distortions, heat capacity experiments show no sign of orbital ordering down to 50 mK and the magnetic susceptibility displays essentially perfect Curie-Weiss behavior with $\theta_{CW}$ = -45.1 K in the range from 15 - 400K \cite{Fritsch2004}.  The possible removal of the ground state orbital degeneracy by random strains was proposed previously \citep{Brossard1976}, but the expected T$^2$ contribution to the specific heat \citep{Ivanov} was not observed experimentally \cite{Fritsch2004}. The orbital degeneracy's contribution to the specific heat and magnetic entropy have been verified experimentally \cite{Fritsch2004}.

In the single ion case, SOC splits the upper orbital triplet into three levels with energies $\Delta_{CF} - 3 \lambda_0$, $\Delta_{CF} -  \lambda_0$, and $\Delta_{CF} + 2 \lambda_0$, where $\Delta_{CF}$  is the $^5$E - $^5$T$_{2}$ splitting and $\lambda_0$ is the SOC constant  \cite{low1960,Wittekoek1973}.  The potentially huge tenfold ground state degeneracy coming from $S=2$ and the orbital doublet is split at second order in the spin-orbit interaction into 5 levels equally separated by $ \lambda = 6 \lambda_0^2/\Delta_{CF}$ (Fig. \ref{Fig1}).  The ground state is a spin and orbitally entangled singlet \cite{Slack69} with the form

\small 
\begin{equation}
\psi _g = \frac{1}{\sqrt{2}}\ket{3z^2{-}r^2}\ket{S_z{=}0}{+}\frac{1}{2} \ket{x^2 - y^2}[\ket{S_z{=}{+}2}{+}\ket{S_z {=}{-}2}].
\nonumber
\end{equation}
\normalsize

 \noindent The first excited state is a spin-orbital triplet, predicted to split Zeeman-like with $g$-factors of $\pm  [1 - 2 (\lambda_0/\Delta_{CF})]$ \cite{low1960}. 

It was suggested by Chen, Balents, and Schnyder \cite{Chen2009a,Chen2009} that the spin and orbitally entangled singlet character of the wavefunction is preserved when the Fe$^{2+}$ ion is incorporated into the FeSc$_2$S$_4$ lattice.   They proposed that the SOL state results from competition between magnetic spin-orbital exchange, which favors a magnetically ordered classical ground state, and on-site SOC, which favors a SOL quantum disordered state.  In the framework of a mean-field next-nearest neighbor (NNN) Kugel-Khomskii-type ``J$_2$/$\lambda$"-model, in which J$_2$ is the NNN exchange constant and $\lambda$ is the excitation energy of single ion Fe$^{2+}$ discussed above, Chen, Balents, and Schnyder predict a quantum phase transition (QPT) at $x_c$ = 1/16  (with $x = J_2 / \lambda$), which separates SOL and ordered phases \cite{Chen2009a,Chen2009}.  The dominance of NNN exchange is demonstrated by the fact that the lowest energy magnetic excitations are found in neutron scattering at the wavevector for a simple N\'eel state ${\bf q} =  \frac{2 \pi }{a} (1, 0, 0 )$ \cite{Krimmel2005} and density functional theory which predicts the ratio of NNN to nearest neighbor exchange to be $\approx$ 37 \cite{Sarkar10a}.  The SOL presumably differs from the ionic limit in that spin and orbital degrees of freedom may be entangled over longer length scales in that a spin on one site becomes entangled with the orbital of another site.  Presumably this length scale diverges as the system approaches the QPT.  The estimated value for $x$ was such as to put FeSc$_2$S$_4$ slightly into the ordered regime \cite{Comment1}.  However, the actual proximity to the QPT and the nature of the ground state has yet to be verified.

Intuitively, one might expect the spectrum of FeSc$_2$S$_4$ in the SOL phase to be similar to that of the single ion Fe$^{2+}$ described above since the SOL results from degeneracies on individual Fe$^{2+}$ and FeSc$_2$S$_4$ breaks no other symmetries aside from those inherent to the crystal.  Chen, Balents, and Schnyder \cite{Chen2009a,Chen2009} predicted that while the first excited state would be a triplet, its energy would be substantially renormalized by exchange.   In the simplest case with only NNN exchange (e.g. the J$_2$/$\lambda$ model) through an expansion in the exchange (valid at $x \ll x_c $) it was shown that the lowest singlet-triplet excitation energy of FeSc$_2$S$_4$ is

\begin{equation}
E (q) = \lambda + 2J_2 \sum_{A} \mathrm{cos}( \bf{q}   \cdot a),
\label{dispersion}
\end{equation}

\noindent where \textbf{a} represents the lattice vectors of the 12 NNN. Such excitations can be probed with optical measurements such as time domain terahertz spectroscopy (TDTS) through the magnetic dipole operator.  Due to negligible momentum of light compared to the lattice scale, we probe the $\bf{q} \rightarrow 0$ limit, reducing Eq. \ref{dispersion} to $E  = \lambda (1+24x)$.  For FeSc$_2$S$_4$, with $x$ of order $x_c$, one expects the excitation energy to be substantially renormalized from the ionic value.

\begin{figure}[tb]
\includegraphics[width=1.0\columnwidth] {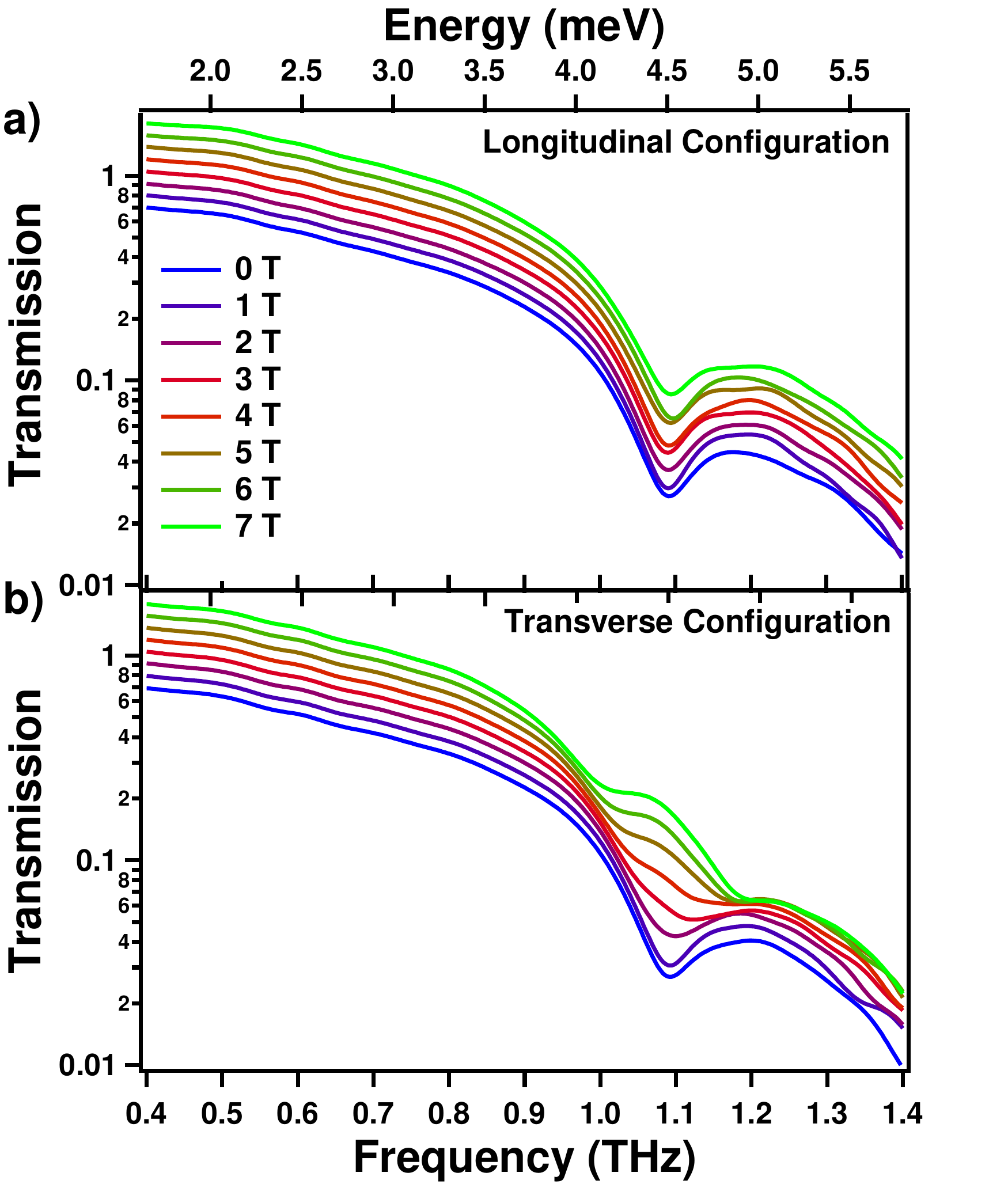}
\caption{(a-b) Field dependence of the transmission magnitude of FeSc$_2$S$_4$ taken at T = 5K for the longitudinal (a) configuration (THz magnetic field $h_{ac} \parallel H_{dc}$) and transverse (b) configuration ($h_{ac} \perp H_{dc}$).  Offsets of 0.05 are included between the curves for clarity.}
\label{Fig2}
\end{figure}

\begin{figure*}[tb]
\includegraphics[width=2.0\columnwidth]{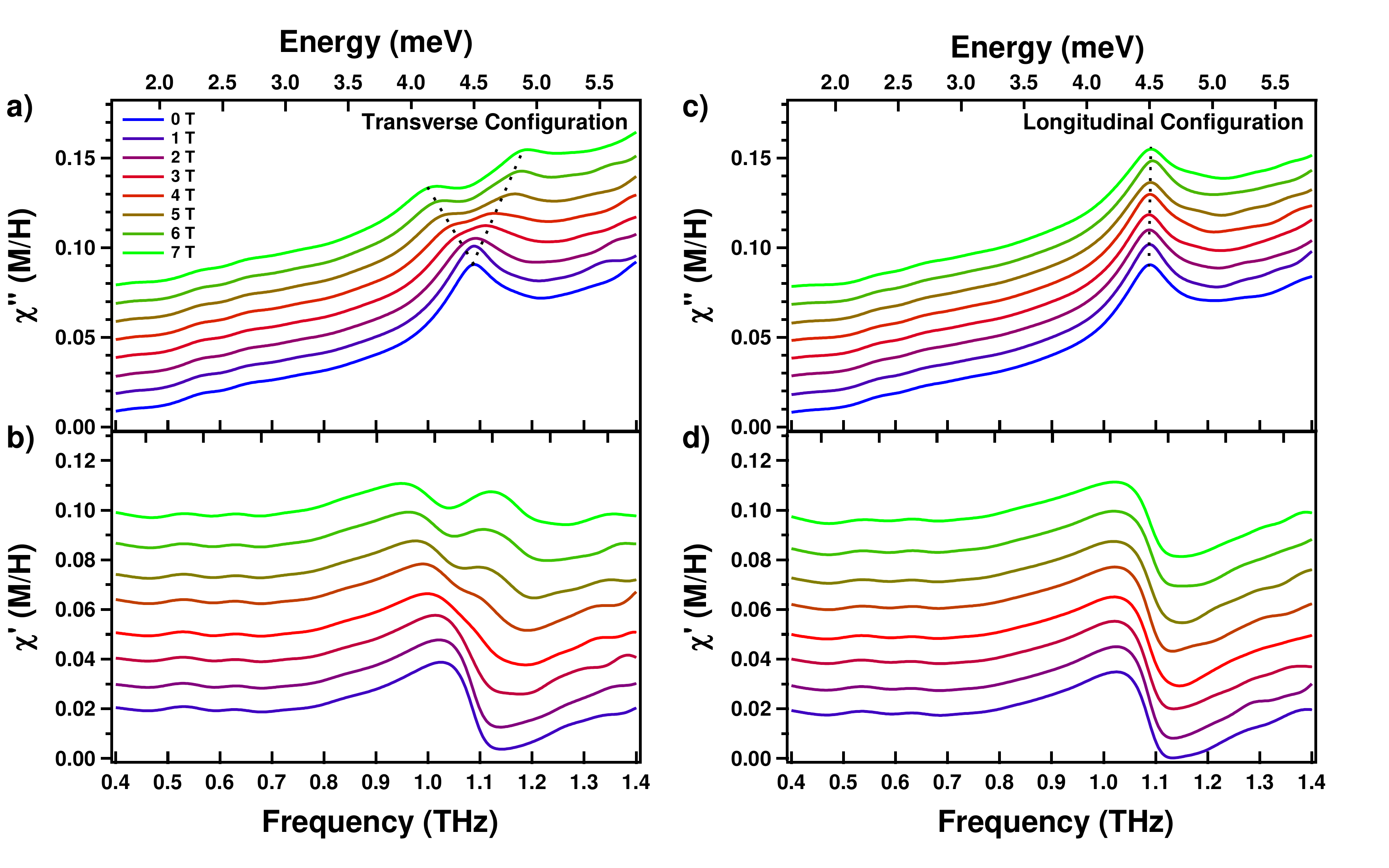}
\caption{Field dependence of the imaginary $\chi''$ and real $\chi'$ parts of the complex ac susceptibility in the transverse (a-b) and longitudinal (c-d) configurations.  All spectra was taken at T = 5K.  The susceptibility is shown in SI units, given by the ratio of the magnetization to applied field.  Dashed lines are guides to the eye.  Offsets of 0.1 are included for clarity.}
\label{Fig3}
\end{figure*}

Dense polycrystalline FeSc$_2$S$_4$ samples were prepared by spark plasma sintering at 1000$^\circ$C from precursor synthesized binary sulfides Sc$_2$S$_3$ and FeS.  Structural, magnetic and optical properties were identical to previous samples prepared by conventional solid state synthesis \cite{Fritsch2004}.  TDTS transmission experiments were performed using a home built spectrometer with applied magnetic fields up to 7 T in the Voigt geometry (light $\bf{k} \perp H_{dc}$).  TDTS is a high resolution method for accurately measuring the electromagnetic response of a sample in the experimentally challenging THz range.  Coupling of the THz fields to magnetic dipole transitions allows access to the frequency dependent \textit{complex} magnetic susceptibility between 100 GHz and 2 THz.  Through the use of a rotating polarizer technique \cite{Morris2012,Comment2}, we measure the sample's response to two polarization directions with respect to $H_{dc}$ simultaneously.  Reflectivity measurements were performed in the mid-infrared frequency range from 1000 to 8000 cm$^{-1}$ using a BRUKER IFS 113v Fourier-transform spectrometer equipped with a He flow cryostat.

Figs. 2a and 2b show the magnitude of the T = 5K transmission coefficient as a function of applied field for the longitudinal (THz magnetic field $h_{ac} \parallel H_{dc}$)  and transverse (THz magnetic field $h_{ac} \perp H_{dc}$) configurations.  In zero field, a sharp absorption develops at 1.08 THz (4.46 meV) below 10K \cite{Mittelstadt2015}.  As shown below, its energy is in reasonable agreement with the predicted singlet-triplet excitation energy of Eq. \ref{dispersion}.  Further evidence for this assignment is found in the field dependence of the transmission.  While no splitting is observed in the longitudinal configuration, the transverse configuration shows a clear splitting into two separate resonances with increasing field, suggesting the presence of distinct selection rules in the system.

Pure spin singlet-triplet excitations are typically forbidden in electron spin resonance measurements due to the parity change between spin singlet and triplet states \cite{Sakai2001} \cite{Huvonen} and are usually only seen in the presence of a Dzyaloshinskii - Moriya (DM) interaction or staggered magnetization arrangements along crystallographic axes.  Chen, Balents, and Schnyder estimate the static DM interactions to be $\approx$ 100 times weaker than both exchange and spin orbit couplings \cite{Chen2009}. In principle, dynamic DM interactions can also weakly allow pure spin singlet-triplet excitations, \cite{Huvonen} but such interactions involve a phonon and are assumed to be even weaker than static contributions. However, neither of these effects are thought to be relevant in FeSc$_2$S$_4$ since the spin-orbital singlet and triplet states belong to the same S $=2$ $^5$E-multiplet and are different from the pure spin states \cite{Ish2015}.  In the single ion case of Fe$^{2+}$ in tetrahedral crystal fields, it has been shown that a similar singlet-triplet excitation is magnetic dipole active with selection rules that agree with the results presented in this work \cite{Slack69} \cite{Slack66a}.  Therefore, we believe the observation of this singlet-triplet excitation is further evidence for the entangled spin-orbital singlet character of the ground state and establishes FeSc$_2$S$_4$ as a SOL.

To better resolve the splitting, the complex ac susceptibility of the sample was calculated from the transmission coefficient \cite{comment3}.  Fig. 3 shows the field dependence of the calculated susceptibility for the transverse (3a-3b) and longitudinal (3c-3d) configurations respectively.  The splitting of the resonance with increasing field is apparent in the transverse configuration.  To extract the peak positions for both configurations the spectra was fit using Lorentzian oscillators and a linear background, which related work proposed derives from a continuum of zone boundary pairs of triplet excitations \cite{Mittelstadt2015}.

\begin{figure}[tb]
\includegraphics[width=1.0\columnwidth]{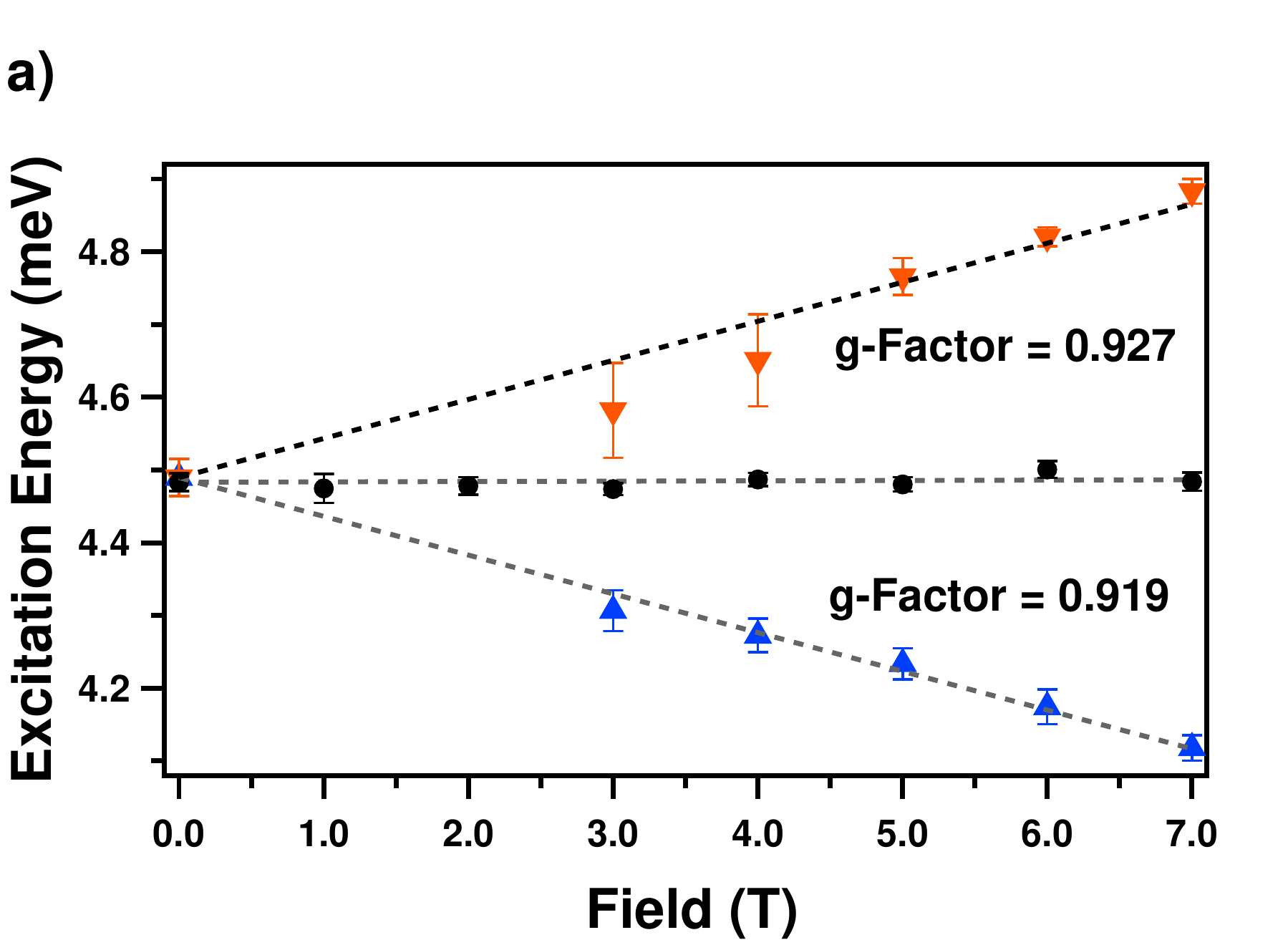}
\includegraphics[width=1.0\columnwidth]{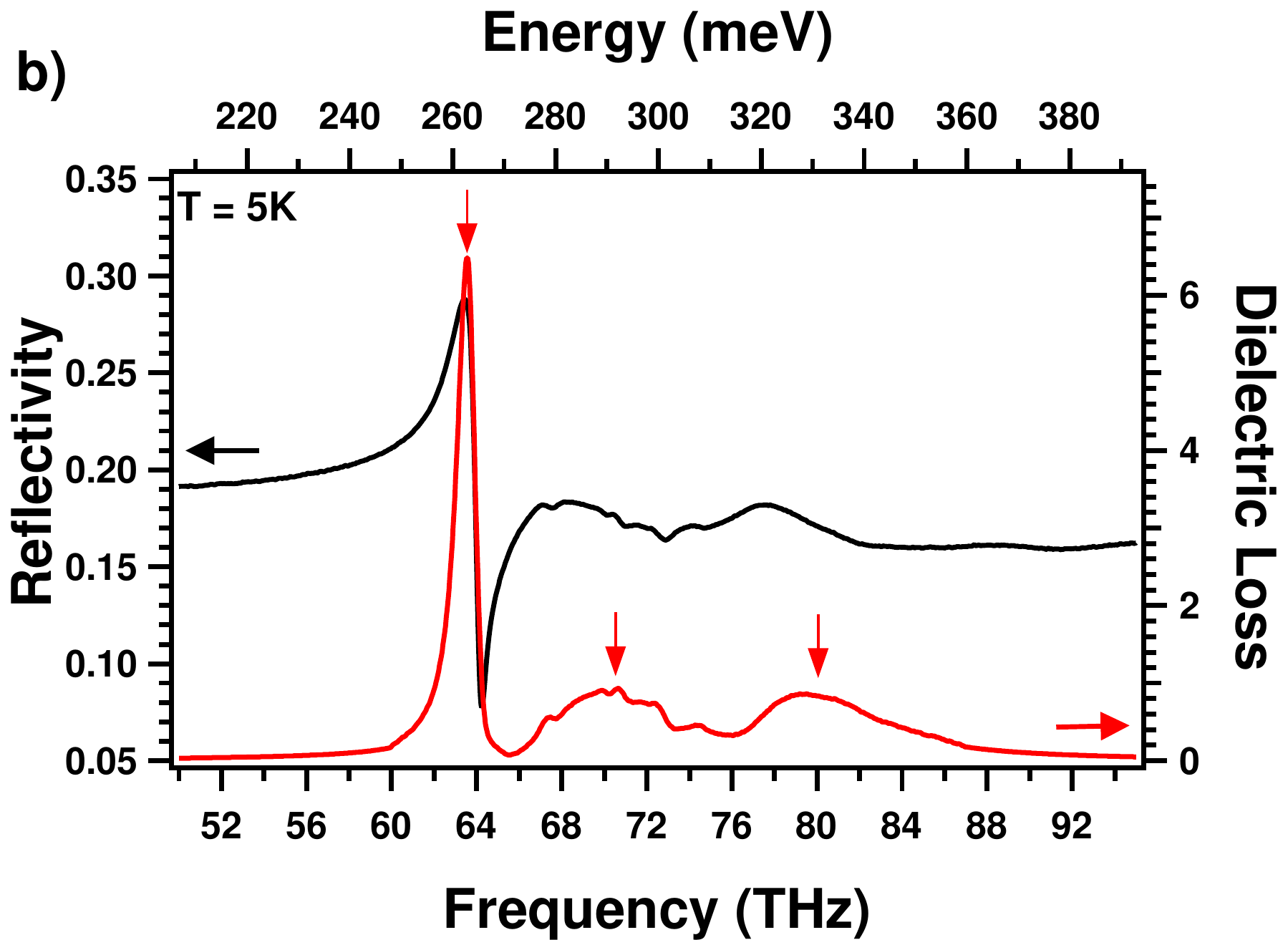}
\caption{(a) Excitation energies for both configurations.  Error bars are based on the quality of the fits. (b) Mid-infrared reflectivity data taken at T = 5K (black, left axis) along with the dielectric loss (red, right axis).  Three features are seen with energies of 63.52 THz (0.262 eV), 70.32 THz (0.290 eV), 80.16 THz (0.331 eV), corresponding to crystal field excitations to the $^5$T$_{2}$ energy levels plus coupling to Jahn-Teller modes of Fe$^{2+}$ in a tetrahedral environment.}
\label{Fig4}
\end{figure}

Fig. \ref{Fig4}a shows the field dependent splitting of the excited state triplet.  Quantitative values for the splittings could not be resolved below 3 T.  Linear fits were performed on each branch to determine the $g$-factors.  Only data between 5T $\leq$ H$_{dc}\leq$ 7T was used in the upper branch fit since low transmission in the high frequency range causes error below 5 T.  We find $g$-factors of 0.93 and -0.92 for the upper and lower branches respectively.

Mid-infrared reflectivity at higher energy was performed in order to determine the crystal field splitting and SOC constant of the sample.  Fig. \ref{Fig4}b shows the 5K MIR reflectivity (black, left axis) spectrum, which displays a number of prominent features.   The excitation energies can be found from the maxima of the dielectric loss (indicated by arrows), which was obtained by using a Kramers Kronig consistent variable dielectric function fitting routine \cite{Kuzmenko2005}.  Although one only expects two optically active excitations from $^5$E to $^5$T$_{2}$ states, additional and shifted absorptions are expected due to strong coupling of the $^5$T$_{2}$ levels to vibrational modes \cite{Slack66a}.  Following the approach of \textcite{Wittekoek1973}, the crystal field splitting, SOC constant, Jahn-Teller coupling mode energies ($E_{JT}$), and coupling constants ($\hbar \omega_{JT}$) can be extracted from the mode energies and intensities.  We determine values of $\Delta_{CF}$ = 71.6 $\pm$ 5 THz (296.1 $\pm$ 20.7 meV), $\lambda_0$ = 2.14 $\pm$ 0.30 THz (8.85 $\pm$ 1.24 meV), $E_{JT}/ \lambda \approx 1.6$, and  $\hbar \omega_{JT} / \lambda \approx 4$.  From these values we can calculate $\lambda$ = 6$\lambda _0 ^2 / \Delta _{CF}$ = 0.38 $\pm$ 0.06 THz (1.57 $\pm$ 0.25 meV). These values correspond closely to values found in other Fe$^{2+}$ tetrahedral compounds \cite{Feiner1982}. 

From our experimental value of $\lambda$ and the value of J$_2$ \cite{Comment1} extracted from the Curie-Weiss constant, we find from Eq. \ref{dispersion} an expected singlet-triplet excitation energy of 1.31 THz (5.42 meV), which is in reasonable agreement with the observed energy of 1.08 THz (4.46 meV).  Additionally, substituting our measured values into the predicted g-factor expression $g =  \pm  [1 - (2 \lambda_0/\Delta_{CF})]$ \cite{low1960} gives expected values of $\pm$ 0.94, which are in excellent agreement with our observed g-factors of 0.93 and -0.92.

With the energy scales characterized, we can work backwards to determine FeSc$_2$S$_4$'s proximity to the QPT in the context of the theory of Chen, Balents, and Schnyder \cite{Chen2009a,Chen2009}.  With an observed excitation energy of 1.08 THz (4.46 meV) and our experimental value for $\lambda$ we can solve Eq. \ref{dispersion} for $x$. Here the implicit assumption is that Eq. \ref{dispersion}, which was considered valid far from the critical point as an expansion in the exchange, is still valid near the QPT for momenta far from the ordering wavevector. We find a value of $x = 0.08$, which puts FeSc$_2$S$_4$ slightly above the predicted $x_c = 1/16$ from mean field theory.  The fact that FeSc$_2$S$_4$ does not order down to the lowest measured temperatures indicates that quantum fluctuations are presumably important in setting $x_c$.   We may use this value of $x$ to make an estimate for J$_2$ of 0.029 THz (0.120 meV), which is about 25\% less than the value inferred from the Curie Weiss constant \cite{Chen2009a,Chen2009,Comment1}.

In this work, we have demonstrated the spin-orbital singlet character of the ground state of FeSc$_2$S$_4$ through the observation of a singlet-triplet excitation.   Its energy is significantly renormalized by the exchange interaction in agreement with the model of Ref. \cite{Chen2009a,Chen2009}.  This system, in close proximity to the QPT,  differs from a simple ensemble of spin-orbit singlet ions through the presence of longer range correlations.   As discussed in Ref. \cite{Chen2009a,Chen2009}, it is believed that the critical regime of this QPT can be described by a Euclidean multicomponent $\Phi^4$ scalar field theory in 4 space-time dimensions.   In such field theories a correlation length can be extracted through the relation $\xi = (h v/E)$ where $E$ is a characteristic energy that vanishes at the QPT and $v$ is a velocity, whose square is a proportionality between space and time derivatives in the effective Lagrangian.  This length can be understood as the scale over which spin and orbital degrees of freedom are entangled.  In the present case, $E \approx 0.17$ meV can be identified with the zone boundary soft gap in neutron scattering \cite{Krimmel2005}.  By inspection of the terms in the action written down in Ref. \cite{Chen2009}, we can identify $v = (a/8 h) \sqrt{\lambda^3/J_2}$ and find $\xi = (\lambda a /8 E) \sqrt{1/x}$.   Using our experimentally determined spin-orbit and exchange parameters we estimate a correlation length of $ \xi/ ( \textbf{a}/2)  \approx 8.2$.   This demonstrates the long-range entangled character of the SOL non-classical ground state in FeSc$_2$S$_4$.

The THz instrumentation and development work at Johns Hopkins was supported by the Gordon an Betty Moore Foundation through grant No. GBMF2628.  N.J.K was supported through the DOE-BES through DE-FG02-08ER46544 and the ARCS Foundation.  In Augsburg this work was supported by the Deutsche Forschungsgemeinschaft via the Transregional Collaborative Research Center TRR 80.   We would like to thank L. Balents, G. Chen, M. V. Eremin, D. Ish, J. Kaplan, O. Tchernyshyov, and A. Turner for helpful conversations.

\bibliography{FeScSBib}

\begin{thebibliography}{25}
\expandafter\ifx\csname natexlab\endcsname\relax\def\natexlab#1{#1}\fi
\expandafter\ifx\csname bibnamefont\endcsname\relax
  \def\bibnamefont#1{#1}\fi
\expandafter\ifx\csname bibfnamefont\endcsname\relax
  \def\bibfnamefont#1{#1}\fi
\expandafter\ifx\csname citenamefont\endcsname\relax
  \def\citenamefont#1{#1}\fi
\expandafter\ifx\csname url\endcsname\relax
  \def\url#1{\texttt{#1}}\fi
\expandafter\ifx\csname urlprefix\endcsname\relax\def\urlprefix{URL }\fi
\providecommand{\bibinfo}[2]{#2}
\providecommand{\eprint}[2][]{\url{#2}}

\bibitem[{\citenamefont{Balents}(2010)}]{Balents2010}
\bibinfo{author}{\bibfnamefont{L.}~\bibnamefont{Balents}},
  \bibinfo{journal}{Nature} \textbf{\bibinfo{volume}{464}},
  \bibinfo{pages}{199} (\bibinfo{year}{2010}).

\bibitem[{\citenamefont{Khaliullin and Maekawa}(2000)}]{Khaliullin2000}
\bibinfo{author}{\bibfnamefont{G.}~\bibnamefont{Khaliullin}} \bibnamefont{and}
  \bibinfo{author}{\bibfnamefont{S.}~\bibnamefont{Maekawa}},
  \bibinfo{journal}{Phys. Rev. Lett.} \textbf{\bibinfo{volume}{85}},
  \bibinfo{pages}{3950} (\bibinfo{year}{2000}).

\bibitem[{\citenamefont{Fritsch et~al.}(2004)\citenamefont{Fritsch, Hemberger,
  B\"uttgen, Scheidt, Krug~von Nidda, Loidl, and Tsurkan}}]{Fritsch2004}
\bibinfo{author}{\bibfnamefont{V.}~\bibnamefont{Fritsch}},
  \bibinfo{author}{\bibfnamefont{J.}~\bibnamefont{Hemberger}},
  \bibinfo{author}{\bibfnamefont{N.}~\bibnamefont{B\"uttgen}},
  \bibinfo{author}{\bibfnamefont{E.-W.} \bibnamefont{Scheidt}},
  \bibinfo{author}{\bibfnamefont{H.-A.} \bibnamefont{Krug~von Nidda}},
  \bibinfo{author}{\bibfnamefont{A.}~\bibnamefont{Loidl}}, \bibnamefont{and}
  \bibinfo{author}{\bibfnamefont{V.}~\bibnamefont{Tsurkan}},
  \bibinfo{journal}{Phys. Rev. Lett.} \textbf{\bibinfo{volume}{92}},
  \bibinfo{pages}{116401} (\bibinfo{year}{2004}).

\bibitem[{\citenamefont{Chen et~al.}(2009{\natexlab{a}})\citenamefont{Chen,
  Balents, and Schnyder}}]{Chen2009a}
\bibinfo{author}{\bibfnamefont{G.}~\bibnamefont{Chen}},
  \bibinfo{author}{\bibfnamefont{L.}~\bibnamefont{Balents}}, \bibnamefont{and}
  \bibinfo{author}{\bibfnamefont{A.~P.} \bibnamefont{Schnyder}},
  \bibinfo{journal}{Phys. Rev. Lett.} \textbf{\bibinfo{volume}{102}},
  \bibinfo{pages}{096406} (\bibinfo{year}{2009}{\natexlab{a}}).

\bibitem[{\citenamefont{Chen et~al.}(2009{\natexlab{b}})\citenamefont{Chen,
  Schnyder, and Balents}}]{Chen2009}
\bibinfo{author}{\bibfnamefont{G.}~\bibnamefont{Chen}},
  \bibinfo{author}{\bibfnamefont{A.~P.} \bibnamefont{Schnyder}},
  \bibnamefont{and} \bibinfo{author}{\bibfnamefont{L.}~\bibnamefont{Balents}},
  \bibinfo{journal}{Phys. Rev. B} \textbf{\bibinfo{volume}{80}},
  \bibinfo{pages}{224409} (\bibinfo{year}{2009}{\natexlab{b}}).

\bibitem[{\citenamefont{Ole\'s}(2012)}]{Ole2012}
\bibinfo{author}{\bibfnamefont{A.~M.} \bibnamefont{Ole\'s}},
  \bibinfo{journal}{Journal of Physics: Condensed Matter}
  \textbf{\bibinfo{volume}{24}}, \bibinfo{pages}{313201}
  (\bibinfo{year}{2012}).

\bibitem[{\citenamefont{Brzezicki et~al.}(2015)\citenamefont{Brzezicki,
  Ole\ifmmode~\acute{s}\else \'{s}\fi{}, and Cuoco}}]{Brzezicki2015}
\bibinfo{author}{\bibfnamefont{W.}~\bibnamefont{Brzezicki}},
  \bibinfo{author}{\bibfnamefont{A.~M.} \bibnamefont{Ole\ifmmode~\acute{s}\else
  \'{s}\fi{}}}, \bibnamefont{and}
  \bibinfo{author}{\bibfnamefont{M.}~\bibnamefont{Cuoco}},
  \bibinfo{journal}{Phys. Rev. X} \textbf{\bibinfo{volume}{5}},
  \bibinfo{pages}{011037} (\bibinfo{year}{2015}).

\bibitem[{\citenamefont{Brossard and Oudet}(1976)}]{Brossard1976}
\bibinfo{author}{\bibfnamefont{P.~G.~L.} \bibnamefont{Brossard}}
  \bibnamefont{and} \bibinfo{author}{\bibfnamefont{H.}~\bibnamefont{Oudet}},
  \bibinfo{journal}{J. Phys. (Paris), Colloq.} \textbf{\bibinfo{volume}{37}}
  (\bibinfo{year}{1976}).

\bibitem[{\citenamefont{Ivanov et~al.}(1983)\citenamefont{Ivanov, Mitrofanov,
  Falkovskaya, and Fishman}}]{Ivanov}
\bibinfo{author}{\bibfnamefont{M.}~\bibnamefont{Ivanov}},
  \bibinfo{author}{\bibfnamefont{V.}~\bibnamefont{Mitrofanov}},
  \bibinfo{author}{\bibfnamefont{L.}~\bibnamefont{Falkovskaya}},
  \bibnamefont{and} \bibinfo{author}{\bibfnamefont{A.}~\bibnamefont{Fishman}},
  \bibinfo{journal}{Journal of Magnetism and Magnetic Materials}
  \textbf{\bibinfo{volume}{36}}, \bibinfo{pages}{26 } (\bibinfo{year}{1983}),
  ISSN \bibinfo{issn}{0304-8853}.

\bibitem[{\citenamefont{Low and Weger}(1960)}]{low1960}
\bibinfo{author}{\bibfnamefont{W.}~\bibnamefont{Low}} \bibnamefont{and}
  \bibinfo{author}{\bibfnamefont{M.}~\bibnamefont{Weger}},
  \bibinfo{journal}{Phys. Rev.} \textbf{\bibinfo{volume}{118}},
  \bibinfo{pages}{1119} (\bibinfo{year}{1960}).

\bibitem[{\citenamefont{Wittekoek et~al.}(1973)\citenamefont{Wittekoek, van
  Stapele, and Wijma}}]{Wittekoek1973}
\bibinfo{author}{\bibfnamefont{S.}~\bibnamefont{Wittekoek}},
  \bibinfo{author}{\bibfnamefont{R.~P.} \bibnamefont{van Stapele}},
  \bibnamefont{and} \bibinfo{author}{\bibfnamefont{A.~W.~J.}
  \bibnamefont{Wijma}}, \bibinfo{journal}{Phys. Rev. B}
  \textbf{\bibinfo{volume}{7}}, \bibinfo{pages}{1667} (\bibinfo{year}{1973}).

\bibitem[{\citenamefont{Slack et~al.}(1969)\citenamefont{Slack, Roberts, and
  Vallin}}]{Slack69}
\bibinfo{author}{\bibfnamefont{G.~A.} \bibnamefont{Slack}},
  \bibinfo{author}{\bibfnamefont{S.}~\bibnamefont{Roberts}}, \bibnamefont{and}
  \bibinfo{author}{\bibfnamefont{J.~T.} \bibnamefont{Vallin}},
  \bibinfo{journal}{Phys. Rev.} \textbf{\bibinfo{volume}{187}},
  \bibinfo{pages}{511} (\bibinfo{year}{1969}).

\bibitem[{\citenamefont{Krimmel et~al.}(2005)\citenamefont{Krimmel, M\"ucksch,
  Tsurkan, Koza, Mutka, and Loidl}}]{Krimmel2005}
\bibinfo{author}{\bibfnamefont{A.}~\bibnamefont{Krimmel}},
  \bibinfo{author}{\bibfnamefont{M.}~\bibnamefont{M\"ucksch}},
  \bibinfo{author}{\bibfnamefont{V.}~\bibnamefont{Tsurkan}},
  \bibinfo{author}{\bibfnamefont{M.~M.} \bibnamefont{Koza}},
  \bibinfo{author}{\bibfnamefont{H.}~\bibnamefont{Mutka}}, \bibnamefont{and}
  \bibinfo{author}{\bibfnamefont{A.}~\bibnamefont{Loidl}},
  \bibinfo{journal}{Phys. Rev. Lett.} \textbf{\bibinfo{volume}{94}},
  \bibinfo{pages}{237402} (\bibinfo{year}{2005}).

\bibitem[{\citenamefont{Sarkar et~al.}(2010)\citenamefont{Sarkar, Maitra,
  Valent{\'\i}, and Saha-Dasgupta}}]{Sarkar10a}
\bibinfo{author}{\bibfnamefont{S.}~\bibnamefont{Sarkar}},
  \bibinfo{author}{\bibfnamefont{T.}~\bibnamefont{Maitra}},
  \bibinfo{author}{\bibfnamefont{R.}~\bibnamefont{Valent{\'\i}}},
  \bibnamefont{and}
  \bibinfo{author}{\bibfnamefont{T.}~\bibnamefont{Saha-Dasgupta}},
  \bibinfo{journal}{Physical Review B} \textbf{\bibinfo{volume}{82}},
  \bibinfo{pages}{041105} (\bibinfo{year}{2010}).

\bibitem[{Com({\natexlab{a}})}]{Comment1}
\bibinfo{note}{Note that in Ref. \cite{Chen2009a} there is numerical mistake in
  the estimate of $J_2$ from the Curie-Weiss temperature. J$_2$ should have
  been estimated to be about 45\% higher e.g. $J_2 = 1.9$ K. This would have
  put the estimated $x$ for FeSc$_2$S$_4$ over the predicted mean-field
  critical value.}

\bibitem[{\citenamefont{Morris et~al.}(2012)\citenamefont{Morris, Aguilar,
  Stier, and Armitage}}]{Morris2012}
\bibinfo{author}{\bibfnamefont{C.~M.} \bibnamefont{Morris}},
  \bibinfo{author}{\bibfnamefont{R.~V.} \bibnamefont{Aguilar}},
  \bibinfo{author}{\bibfnamefont{A.~V.} \bibnamefont{Stier}}, \bibnamefont{and}
  \bibinfo{author}{\bibfnamefont{N.~P.} \bibnamefont{Armitage}},
  \bibinfo{journal}{Optics Express} \textbf{\bibinfo{volume}{20}}
  (\bibinfo{year}{2012}).

\bibitem[{Com({\natexlab{b}})}]{Comment2}
\bibinfo{note}{The setup differed slightly from that in Ref. \cite{Morris2012}.
  Polarizer P1 was set to 45 degrees from the vertical and RP and P2 were
  placed after OAP 4. P3 was not needed. For a diagonal transmission matrix
  $T$, the in- and out-of-phase lockin outputs in this configuration give a
  simultaneous measure of $T_{xx}$ and $T_{yy}$. For DC magnetic field
  polarized in the $x$ direction this gives the longitudinal and transverse
  responses simultaneously.}

\bibitem[{\citenamefont{Mittelst\"adt et~al.}(2015)\citenamefont{Mittelst\"adt,
  Schmidt, Wang, Mayr, Tsurkan, Lunkenheimer, Ish, Balents, Deisenhofer, and
  Loidl}}]{Mittelstadt2015}
\bibinfo{author}{\bibfnamefont{L.}~\bibnamefont{Mittelst\"adt}},
  \bibinfo{author}{\bibfnamefont{M.}~\bibnamefont{Schmidt}},
  \bibinfo{author}{\bibfnamefont{Z.}~\bibnamefont{Wang}},
  \bibinfo{author}{\bibfnamefont{F.}~\bibnamefont{Mayr}},
  \bibinfo{author}{\bibfnamefont{V.}~\bibnamefont{Tsurkan}},
  \bibinfo{author}{\bibfnamefont{P.}~\bibnamefont{Lunkenheimer}},
  \bibinfo{author}{\bibfnamefont{D.}~\bibnamefont{Ish}},
  \bibinfo{author}{\bibfnamefont{L.}~\bibnamefont{Balents}},
  \bibinfo{author}{\bibfnamefont{J.}~\bibnamefont{Deisenhofer}},
  \bibnamefont{and} \bibinfo{author}{\bibfnamefont{A.}~\bibnamefont{Loidl}},
  \bibinfo{journal}{Phys. Rev. B} \textbf{\bibinfo{volume}{91}},
  \bibinfo{pages}{125112} (\bibinfo{year}{2015}).

\bibitem[{\citenamefont{Sakai et~al.}(2001)\citenamefont{Sakai, Cepas, and
  Ziman}}]{Sakai2001}
\bibinfo{author}{\bibfnamefont{T.}~\bibnamefont{Sakai}},
  \bibinfo{author}{\bibfnamefont{O.}~\bibnamefont{Cepas}}, \bibnamefont{and}
  \bibinfo{author}{\bibfnamefont{T.}~\bibnamefont{Ziman}},
  \bibinfo{journal}{Physica B: Condensed Matter}
  \textbf{\bibinfo{volume}{294-295}} (\bibinfo{year}{2001}).

\bibitem[{\citenamefont{H{\"u}vonen}(2008)}]{Huvonen}
\bibinfo{author}{\bibfnamefont{D.}~\bibnamefont{H{\"u}vonen}}, Ph.D. thesis,
  \bibinfo{school}{National Institute of Chemical Physics and Biophysics,
  Tallinn, Estonia} (\bibinfo{year}{2008}).

\bibitem[{\citenamefont{Ish and Balents}(2015)}]{Ish2015}
\bibinfo{author}{\bibfnamefont{D.}~\bibnamefont{Ish}} \bibnamefont{and}
  \bibinfo{author}{\bibfnamefont{L.}~\bibnamefont{Balents}},
  \bibinfo{journal}{Phys. Rev. B} \textbf{\bibinfo{volume}{92}},
  \bibinfo{pages}{094413} (\bibinfo{year}{2015}).

\bibitem[{\citenamefont{Slack et~al.}(1966)\citenamefont{Slack, Ham, and
  Chrenko}}]{Slack66a}
\bibinfo{author}{\bibfnamefont{G.~A.} \bibnamefont{Slack}},
  \bibinfo{author}{\bibfnamefont{F.~S.} \bibnamefont{Ham}}, \bibnamefont{and}
  \bibinfo{author}{\bibfnamefont{R.~M.} \bibnamefont{Chrenko}},
  \bibinfo{journal}{Phys. Rev.} \textbf{\bibinfo{volume}{152}},
  \bibinfo{pages}{376} (\bibinfo{year}{1966}).

\bibitem[{com()}]{comment3}
\bibinfo{note}{The complex transmission of the sample is given by the relation
  $\widetilde{T}( \omega) = \frac{4 \widetilde{Z}}{{(\widetilde{Z}+1)}^2} e^{i
  \omega d (\tilde{n}-1) / c }$, where $\widetilde{Z} =
  \sqrt{\frac{\mu}{\epsilon}}$ is the complex impedance, $d$ is the sample
  thickness, and $\tilde{n} = \sqrt{\epsilon \mu}$ is the complex index of
  refraction. By assuming the magnetic susceptibility, $\tilde{\chi} $, is
  small compared to the dielectric contribution, we can approximate $\tilde{n}
  \approx n(1 + \frac{\tilde{\chi} }{2})$ and $\widetilde{Z} \approx
  \frac{1}{n}$. The complex ac magnetic susceptibility is then given by
  $\tilde{\chi} = \frac{2}{n}[\frac{ic}{\omega d}[ln(\frac{4n}{(n+1)^2}) -
  ln(\tilde{T}(\omega))] + 1 - n]$}.

\bibitem[{\citenamefont{Kuzmenko}(2005)}]{Kuzmenko2005}
\bibinfo{author}{\bibfnamefont{A.~B.} \bibnamefont{Kuzmenko}},
  \bibinfo{journal}{Review of Scientific Instruments}
  \textbf{\bibinfo{volume}{76}}, \bibinfo{eid}{083108} (\bibinfo{year}{2005}).

\bibitem[{\citenamefont{Feiner}(1982)}]{Feiner1982}
\bibinfo{author}{\bibfnamefont{L.~F.} \bibnamefont{Feiner}},
  \bibinfo{journal}{J. Phys. C.} \textbf{\bibinfo{volume}{15}}
  (\bibinfo{year}{1982}).

\end{thebibliography}

\end{document}